\documentclass[10pt, paper=a4, UKenglish]{article}
\usepackage{graphicx}
\def\Title#1{\begin{center} {\Large #1 } \end{center}}
\def\Author#1{\begin{center}{ \sc #1} \end{center}}
\def\Address#1{\begin{center}{ \it #1} \end{center}}

\newcommand\pubblock{\rightline{\begin{tabular}{l} Proceedings of the CTD/WIT 2019\\ \CTDpubnumber\\ \pubnumber\\
         June 21, 2019  \end{tabular}}}

\newenvironment{Abstract}{\begin{quotation} \begin{center} 
             \large ABSTRACT \end{center}\bigskip 
      \begin{center}\begin{large}}{\end{large}\end{center} \end{quotation}}

\newenvironment{Presented}{\begin{quotation} \begin{center} 
             PRESENTED AT\end{center}\bigskip 
      \begin{center}\begin{large}}{\end{large}\end{center} \end{quotation}}

\def\Acknowledgements{\bigskip  \bigskip \begin{center} \begin{large}
      \bf ACKNOWLEDGEMENTS \end{large}\end{center}}

\def\beq{\begin{equation}}
\def\eeq#1{\label{#1}\end{equation}}
\def\eeqn{\end{equation}}

\def\beqa{\begin{eqnarray}}
\def\eeqa#1{\label{#1}\end{eqnarray}}
\def\eeqan{\end{eqnarray}}

\let\bar=\overbar

\def\Dslash{\not{\hbox{\kern-4pt $D$}}}
\def\dslash{\not{\hbox{\kern-2pt $\del$}}}

\def\msb{{\bar{\ssstyle M \kern -1pt S}}}

\newcommand{\lessSpace}{-3pt}
\newcommand{\imageSize}{0.39\textwidth}

\usepackage{color}
\usepackage{lineno}
\usepackage[subrefformat=parens,labelformat=parens]{subfig}
\usepackage{hyperref}
\usepackage{url}
\usepackage{amsmath}
\usepackage{amssymb}

\newcommand\pubnumber{ATL-PHYS-PROC-2019-048}
\newcommand\CTDpubnumber{PROC-CTD19-035}

\def\affiliation{
On behalf of the ATLAS collaboration, \\
Department of Physics \\
Lancaster University, United Kingdom}

\newcommand{\conference}{Connecting the Dots and Workshop on Intelligent Trackers (CTD/WIT 2019)\\
Instituto de F\'isica Corpuscular (IFIC), Valencia, Spain\\ 
April 2-5, 2019}

\usepackage{fancyhdr}
\definecolor{mygrey}{RGB}{105,105,105}
\begin{document}


\large
\begin{titlepage}
\pubblock

\vfill
\Title{Development of ATLAS Primary Vertex Reconstruction for LHC Run 3}
\vfill

\Author{Izaac Sanderswood}
\Address{\affiliation}
\vfill

\begin{Abstract}
  Increasing luminosity at the Large Hadron Collider (LHC) poses a challenge for primary vertex reconstruction in the ATLAS experiment. A rate of 70 or more inelastic proton-proton collisions per beam crossing was observed during the recently-completed Run 2 and even higher vertex density, or pile-up, is expected in Run 3. To meet this challenge, ATLAS has developed new tools: a Gaussian track density seed finder and an adaptive multi-vertex finder. The former constructs a simple but powerful analytic model of the track density along the beam axis to locate candidate vertices, and the latter applies a global approach to vertex finding and fitting, allowing vertices to compete for nearby tracks. These proceedings document the strategy, optimization and preliminary performance of this new vertex reconstruction software, highlighting improvements in vertex finding efficiency, purity and spatial resolution under Run 3 pile-up conditions.
\end{Abstract}

\vfill

\begin{Presented}
\conference
\end{Presented}
\vfill
\begin{center}
Copyright 2019 CERN for the benefit of the ATLAS Collaboration. CC-BY-4.0 license.
\end{center}
\end{titlepage}
\def\thefootnote{\fnsymbol{footnote}}
\setcounter{footnote}{0}
%

\normalsize 


\vspace{\lessSpace}
\section{Introduction}
\label{sec:Introduction}
Precise reconstruction of primary vertices is paramount in the ATLAS physics programme. Primary vertices are the locations of inelastic proton-proton ($pp$) interactions, the fundamental origins of all reconstructed objects used in physics analyses. With ATLAS experiencing an ever-increasing luminosity, new and improved tools are required to ensure optimal reconstruction. Two such tools, the adaptive multi-vertex fitter (AMVF) \cite{Piacquadio:2008zzb,Piacquadio:1243771} and Gaussian  track density seed finder (GS) are summarised here, with a more detailed description given in \cite{ATLAS:PubNote}. The AMVF will replace the previously used iterative vertex finder (IVF).

Section \ref{sec:PrimaryVertexReconstruction} gives a qualitative description of the differences between the new and old vertex reconstruction strategies. Section \ref{sec:PerformanceComparison} compares the performance of the new and old vertex reconstruction strategies.


\section{Primary vertex reconstruction}
\label{sec:PrimaryVertexReconstruction}
In the ATLAS software framework, primary vertex reconstruction is handled by tools called vertex finders. In brief, primary vertex finders are used to reconstruct a set of primary vertices from a given set of reconstructed tracks. Vertex reconstruction is fundamentally comprised of two components:
\begin{itemize}
	\item vertex finding: the association of tracks to a particular vertex seed
	\item vertex fitting: the reconstruction of the vertex position along with its covariance matrix and estimation of the quality of fit.
\end{itemize}
Vertex finders are largely modular, with different interfaces for track selection, seed finders, vertex fitters, deterministic annealing schedules, impact point estimators, etc. This allows for some degree of portability, though some vertex finders are specific implementations of specific vertex fitters. 

\begin{figure}[htbp]
\begin{center}
\subfloat[]{
  \label{fig:IVF_sketch}
  \includegraphics[width=\imageSize]{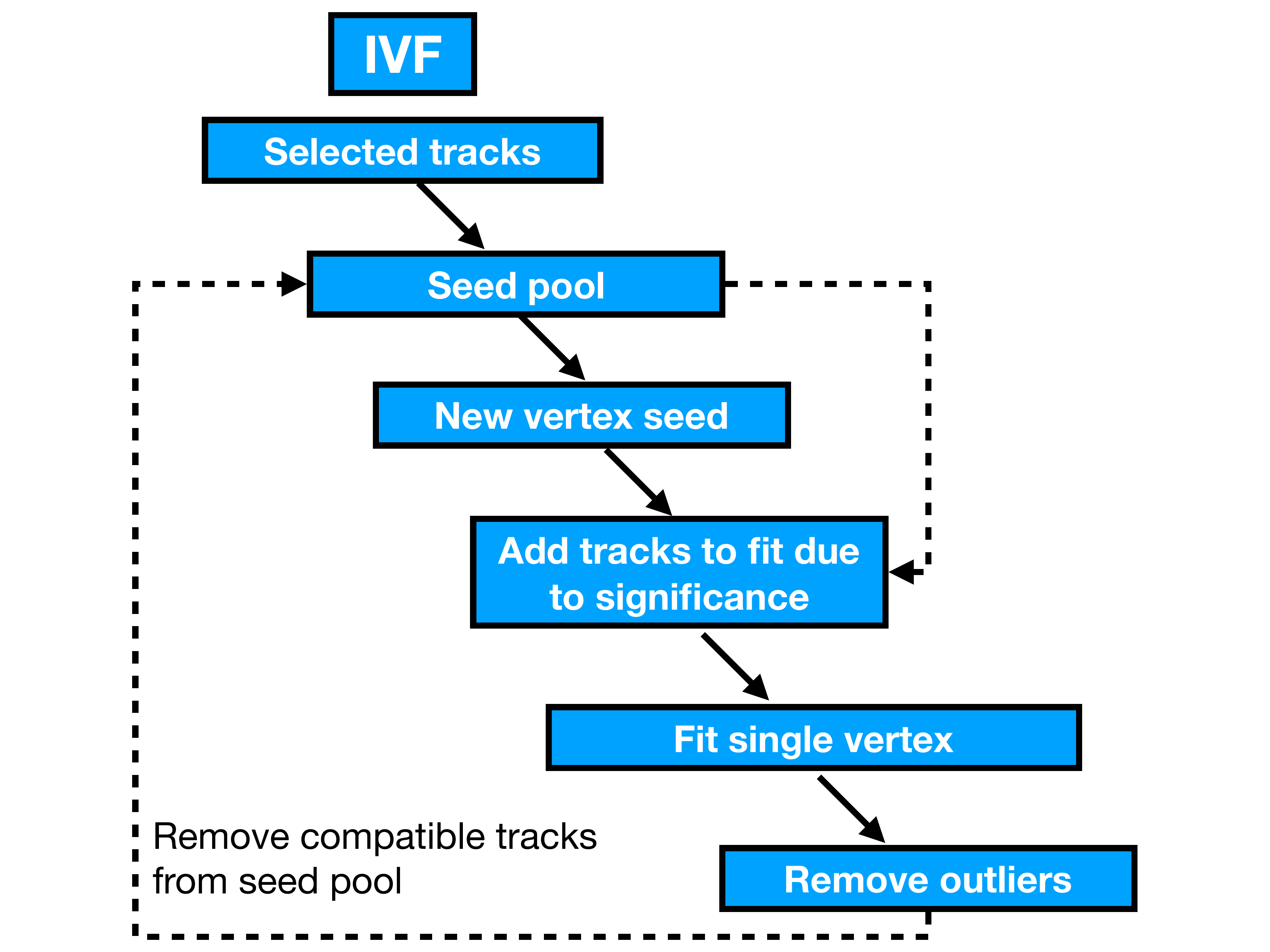}
}
\subfloat[]{
  \label{fig:AMVF_sketch}
  \includegraphics[width=\imageSize]{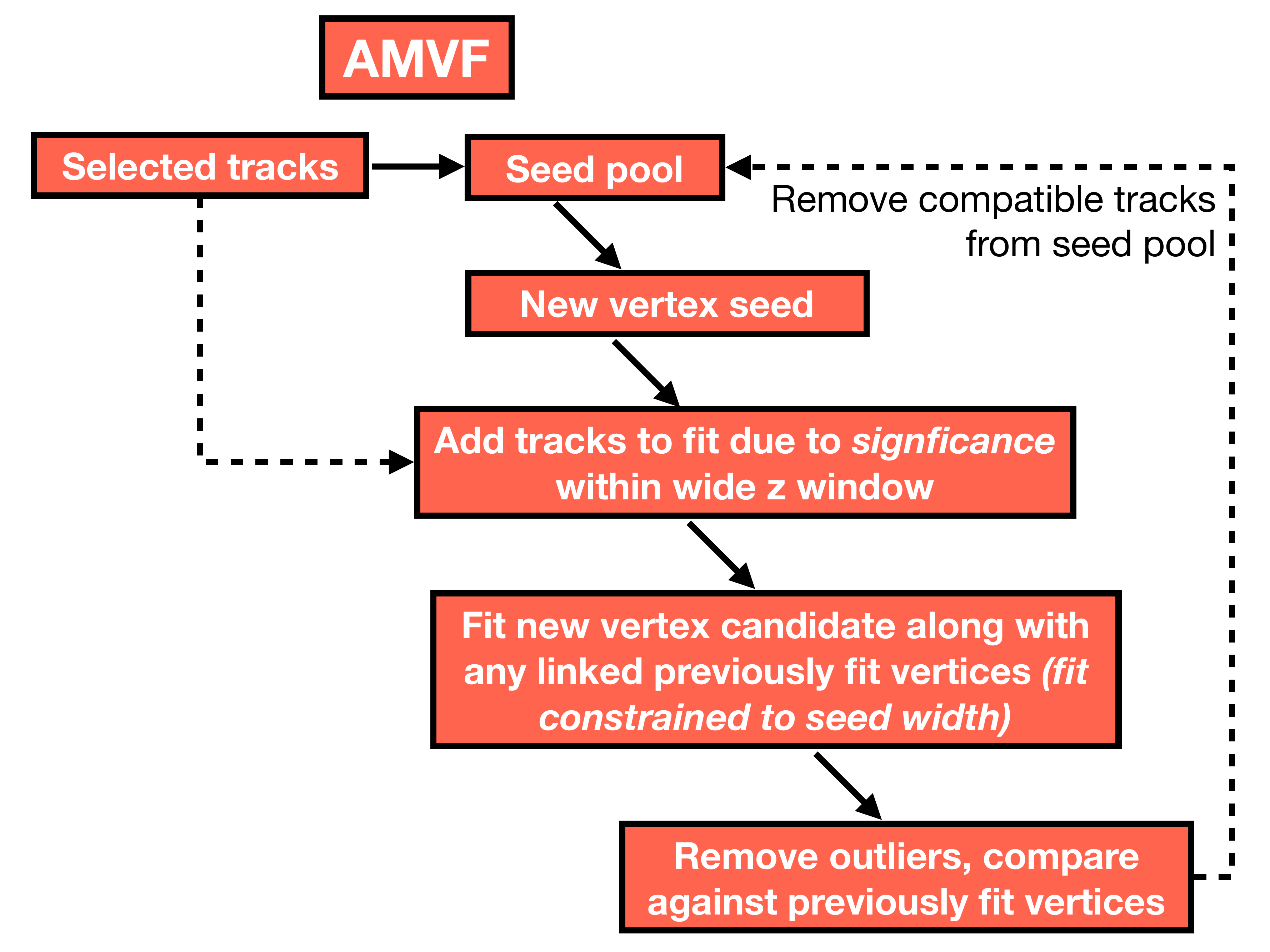}
}
\caption{A simple diagram showing the logic of \protect\subref{fig:IVF_sketch} the IVF and \protect\subref{fig:AMVF_sketch} the AMVF.}
\label{fig:FlowCharts}
\end{center}
\end{figure}
\vspace{\lessSpace}\subsection{AMVF vs IVF}
\label{sec:AMVFvsIVF}
The AMVF and IVF broadly follow the same procedure for vertex reconstruction with some key differences. This section outlines the similarities and differences between the two strategies, though detailed descriptions can be found in Ref. \cite{ATLAS:PubNote}. Flow charts summarising the different strategies are shown in Figure \ref{fig:FlowCharts}. Both vertex finders use implementations of the adaptive estimator \cite{Fruhwirth:2007hz} to calculate the weights of the tracks during the vertex fit. They both use deterministic annealing to progressively de-weight outlier tracks. The AMVF uses a true multi-vertex fitter, meaning tracks can have weights to multiple vertices. The IVF uses a single vertex fitter. Both fit vertices iteratively, one after the other. In the case of the AMVF, if any previously fit vertices share tracks to the vertex candidate currently being fit, these are all fit simultaneously. This means that for the AMVF vertex positions can change and tracks can be reassigned as new candidates are added to the fit (i.e. using a global fit in regions where it is required). The AMVF uses the global track pool to select tracks for new candidates, whereas the IVF only uses tracks not compatible to any previously fit vertices. The iterative finder adds tracks to the fit using a loose significance calculation ($\mathrm{distance}/\sigma_\mathrm{distance}$), whereas the AMVF now uses a new significance cut within a 3 mm wide $z$ window.

The AMVF was developed before ATLAS data-taking started, but the IVF has been used for the duration of data taking. In studies for ATLAS Run 4, the performance of the IVF was shown to degrade significantly with increasing pile-up. The AMVF was shown to have much better and pile-up independent performance, especially after some tuning. Based on these studies it was decided to investigate the prospects of the AMVF for ATLAS Run 3, and whether the AMVF could be further developed and optimised for expected Run 3 conditions.

\vspace{\lessSpace}\subsection{Gaussian track density seed finder}
\label{sec:GS}
A vertex seed is the most likely position of a new vertex candidate. It is determined using tracks not compatible to previously fit vertices using a seed finder. ATLAS previously used a ``fraction of sample mode with weights'' mode finder. The AMVF now uses the newly developed Gaussian Track Density Seed Finder (GS) where seed finding weights are calculated using a longitudinal Gaussian function with a transverse Gaussian function acting as an independent quality control. The key advantages of this method are that it is simpler and accounts for the track uncertainties. Another advantage of this analytic method is that it allows exploitation of the seed width as a longitudinal constraint in the vertex fit. The GS is described in much more detail in Ref. \cite{ATLAS:PubNote}.

\vspace{\lessSpace}\subsection{Simulated data samples}
\label{sec:SimulatedDataSamples}
The simulated data used in these proceedings are generated at centre-of-mass energy $\sqrt{s} = 13$ TeV. Each simulated event contains a single $pp$ interaction involving a large-momentum-transfer process referred to as the ``hard-scatter'' (HS), overlaid with a Poisson-distributed number of minimum-bias inelastic $pp$ interactions (pile-up). Two very different topologies are generated: top-quark pair production ($t\bar{t}$) and a Higgs boson produced via vector boson fusion, decaying into undetected particles ($\mathrm{VBF}\, H \to4\nu$). The production methods are described in Ref. \cite{ATLAS:PubNote}. The simulated data samples have an average mean pile-up of $\langle\mu\rangle$~=~60, with mean pile-up ranging from around $40<\mu<80$.

\vspace{\lessSpace}
\section{Performance comparison}
\label{sec:PerformanceComparison}

This section compares the performance of old and new vertex reconstruction strategies. Performance is compared using measurements of vertex reconstruction quality, reconstruction and identification efficiencies, transverse and longitudinal resolutions, and track association. Local pile-up density is defined as the number of true $pp$ interactions within a 2 mm window around the true signal interaction.

\vspace{\lessSpace}\subsection{Vertex quality}
\label{sec:VertexReconstructionQuality}
\begin{figure}[!htb]
  \centering
  \subfloat[Vertex classifications per event]{
  	\label{fig:PUVertexClassification}
  	\includegraphics[width=\imageSize]{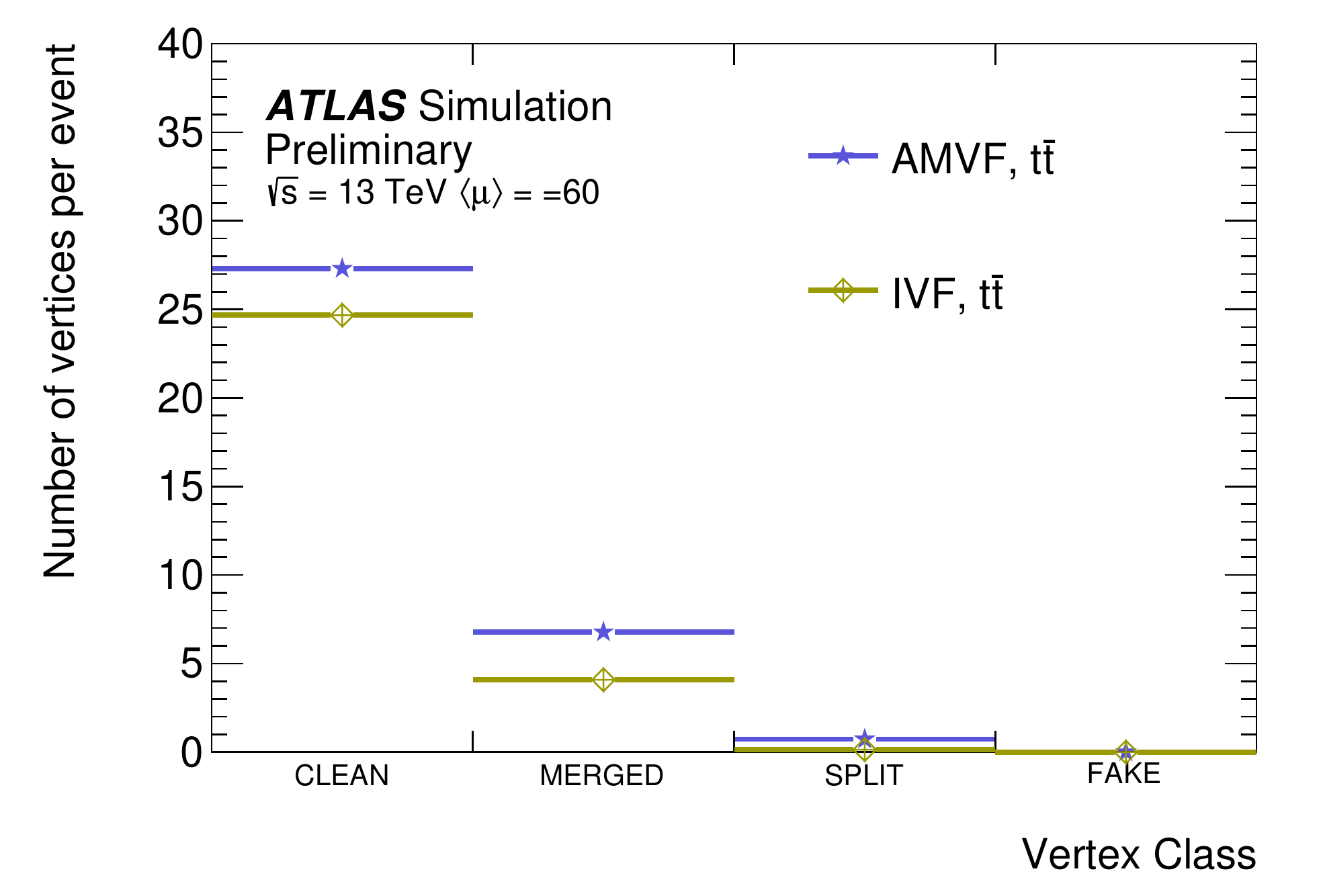}
  }
	\qquad 
  \subfloat[Hard Scatter Vertex Classifications]{
  	\label{fig:HSVertexClassification}
  	\includegraphics[width=\imageSize]{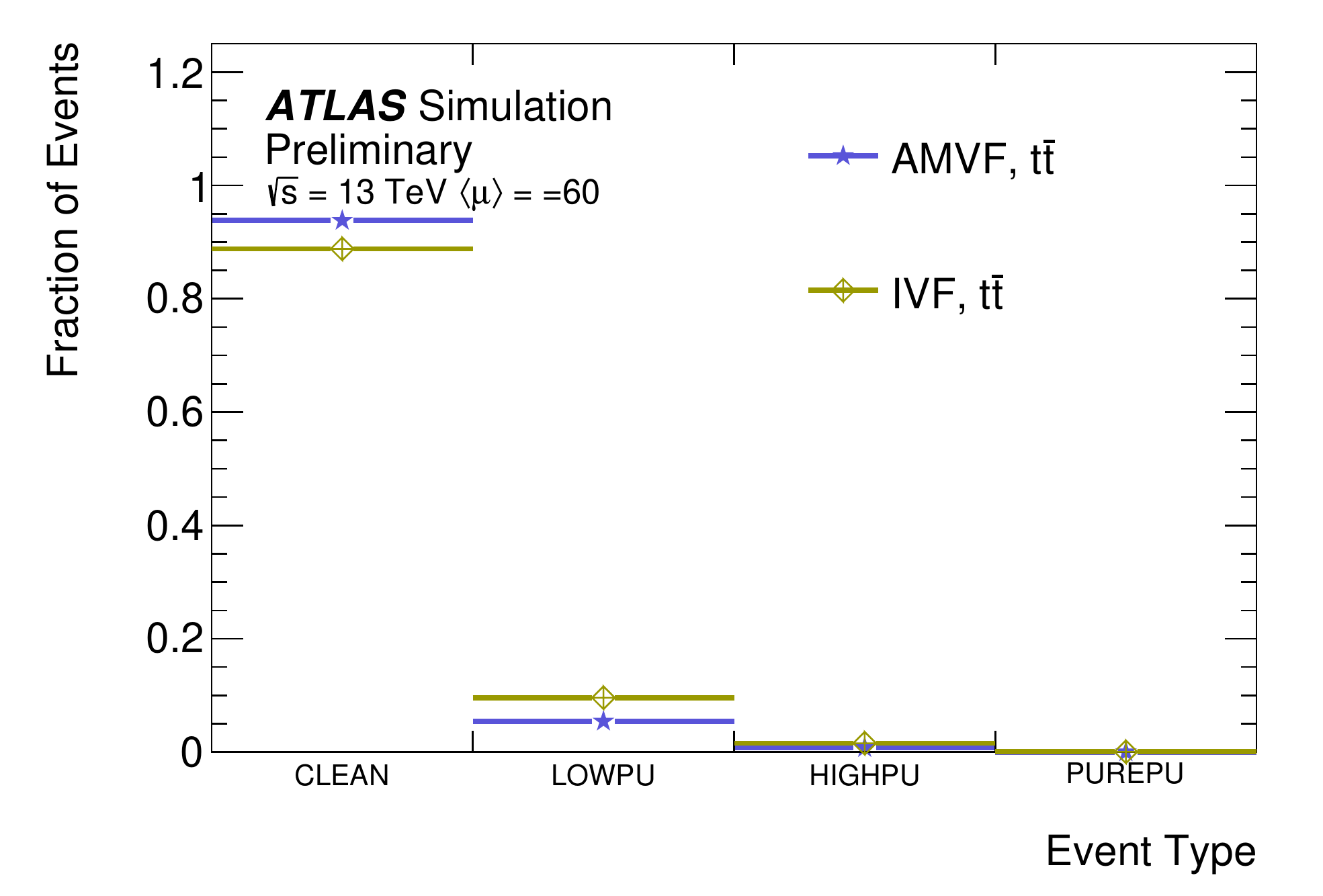}
  	}
  \caption{Comparison between AMVF and IVF vertex quality. Figure \protect\subref*{fig:PUVertexClassification} shows the quality of pile-up vertices. Figure \protect\subref*{fig:HSVertexClassification} shows the quality of the vertex matched to the HS \protect\cite{ATLAS:PubNote}.}
  \label{fig:vertexClassification}
\end{figure}
Figure \subref*{fig:PUVertexClassification} shows the number of reconstructed vertices per event for the different quality grades, for the IVF and AMVF. The quality of the reconstructed vertex matched to the true hard scatter vertex is shown in Figure \subref*{fig:HSVertexClassification}. These grades quantify the correct assignment of tracks to a reconstructed vertex, as well as the pile-up contamination of the HS vertex. CLEAN vertices require at least $70\%$ track weight originating from a single simulated $pp$ interaction. The categories are described in detail in Ref. \cite{ATLAS:PubNote}. In Figure \subref*{fig:PUVertexClassification}, an increase in the number of reconstructed vertices is seen with the AMVF compared to the IVF. Most of these additional vertices are graded as CLEAN/MATCHED and MERGED. Figure \subref*{fig:HSVertexClassification} shows that the AMVF demonstrates a 5\% increase in the number of events graded as CLEAN compared to the IVF.   

\vspace{\lessSpace}\subsection{Hard scatter vertex reconstruction performance}
\label{sec:HardScatterVertexReconstructionAndSelectionEfficiency}
\begin{figure}[!htb]
  \centering
  \subfloat[$t\bar{t}$]{
  	\label{fig:ttbarRecoEff}
  	\includegraphics[width=\imageSize]{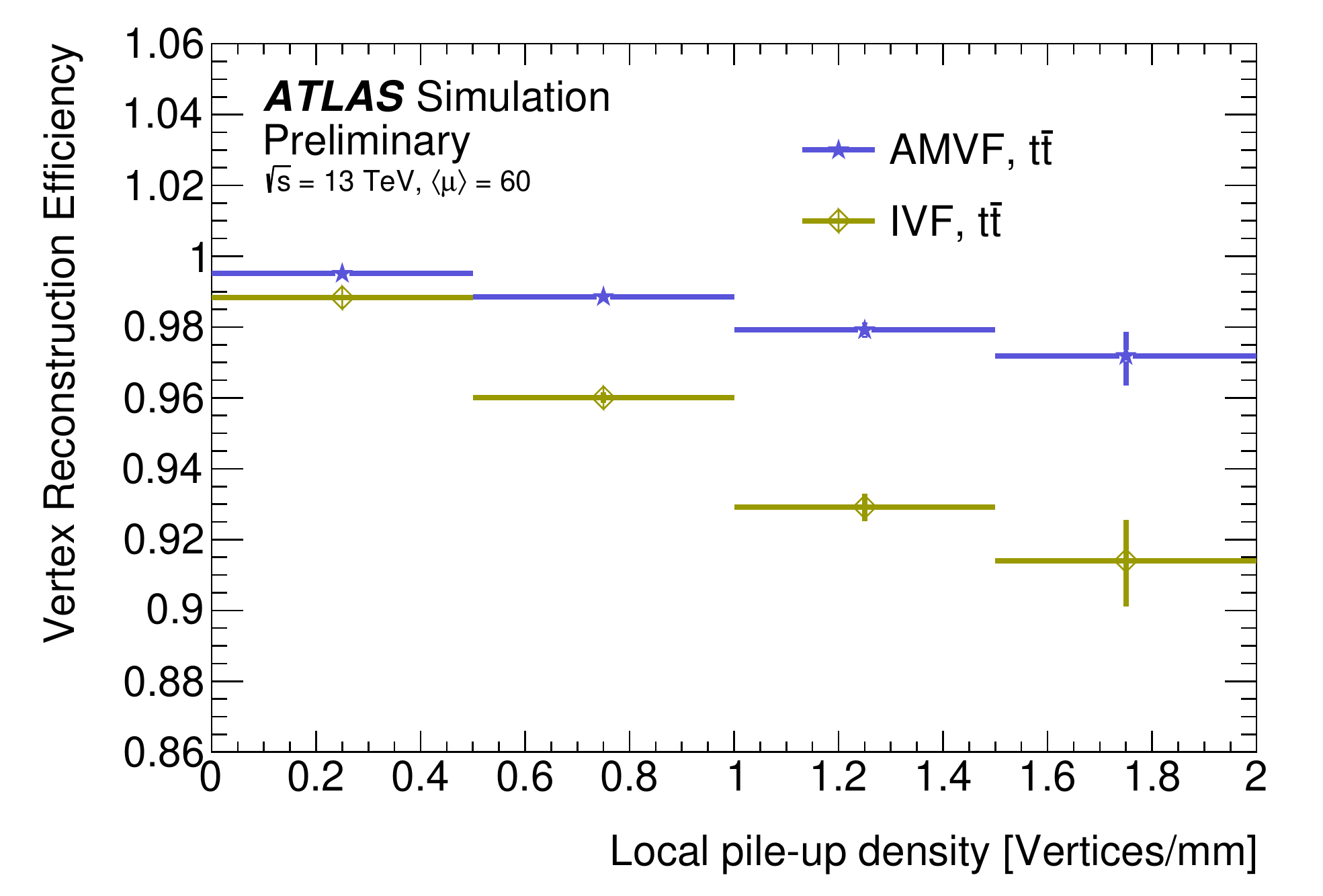}
  }
	\qquad 
  \subfloat[$\mathrm{VBF}\, H\to4\nu$]{
	\label{fig:hinvRecoEff}  
  	\includegraphics[width=\imageSize]{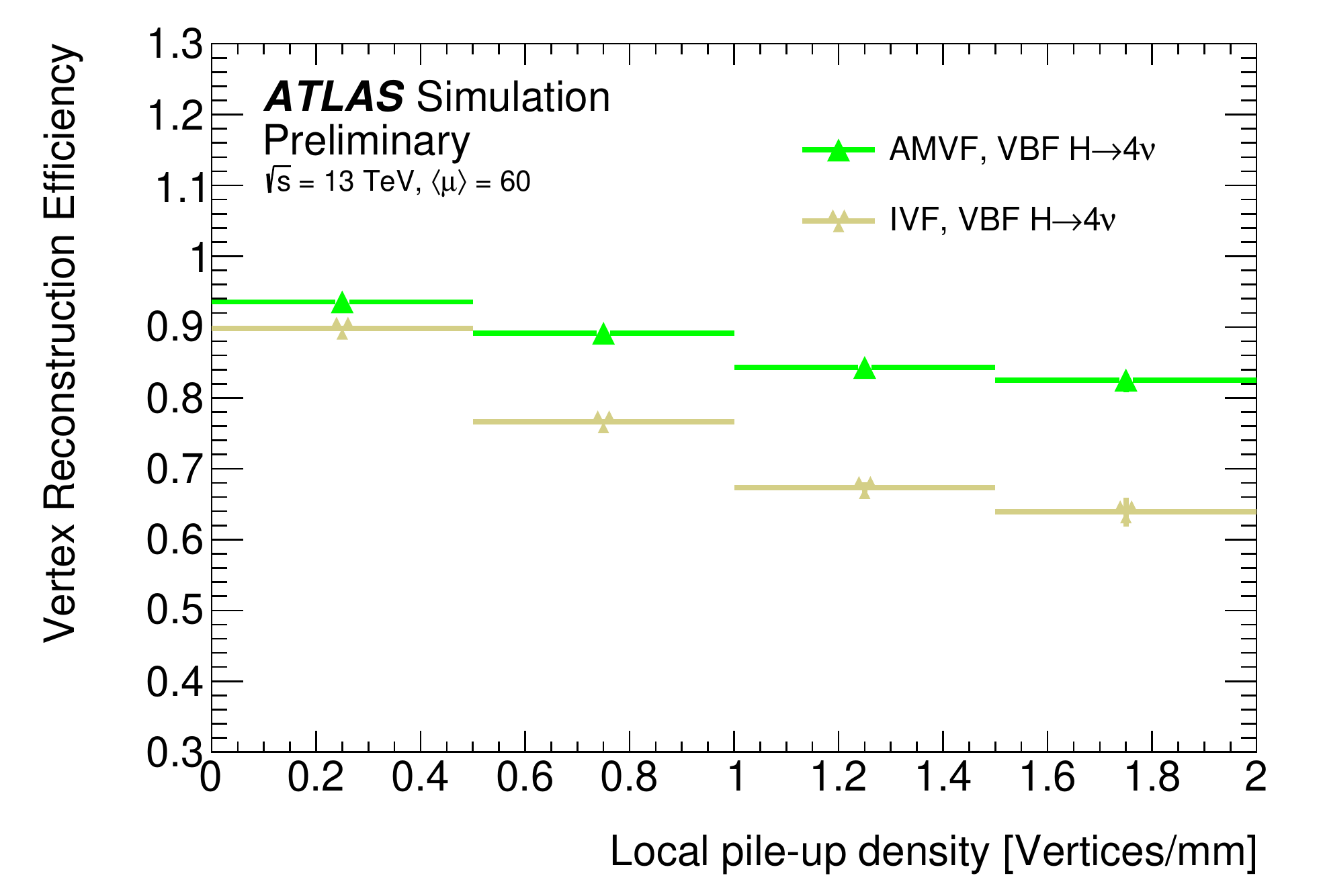}
  }
  \caption{Comparison between AMVF and IVF HS vertex reconstruction efficiencies, as a function of local pile-up density. Figure \protect\subref*{fig:ttbarRecoEff} (Figure \protect\subref*{fig:hinvRecoEff}) shows performance for simulated $t\bar{t}$ ($\mathrm{VBF}\, H\to4\nu$) \protect\cite{ATLAS:PubNote}.}
  \label{fig:recoEfficiencies}
\end{figure}
\begin{figure}[!htb]
  \centering
  \subfloat[Hard scatter vertex selection efficiency]{
  	\label{fig:ttbarIdEff}
  	\includegraphics[width=\imageSize]{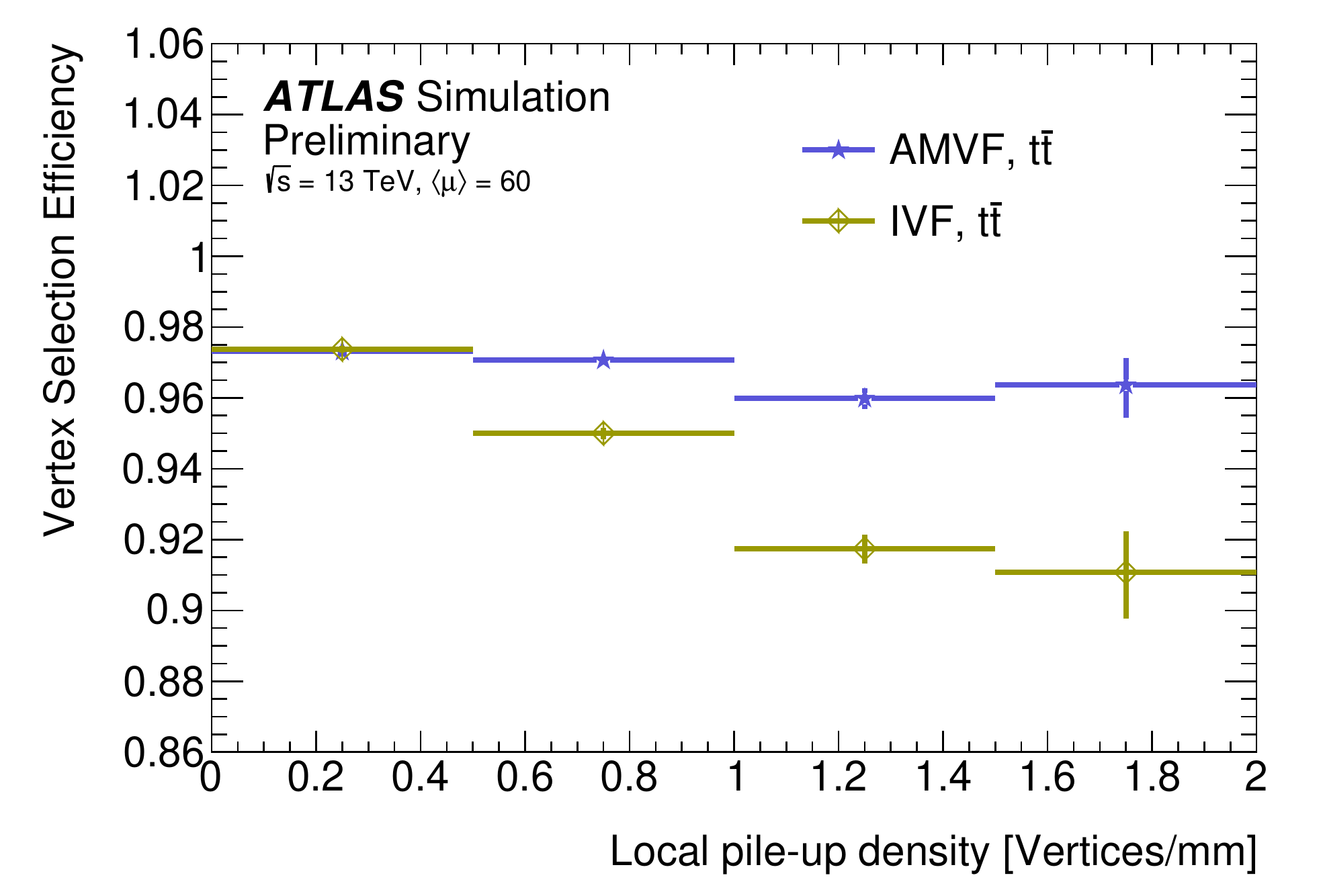}
  }
	\qquad 
  \subfloat[Hard scatter vertex selection efficiency]{
    	\label{fig:hinvIdEff}
    	\includegraphics[width=\imageSize]{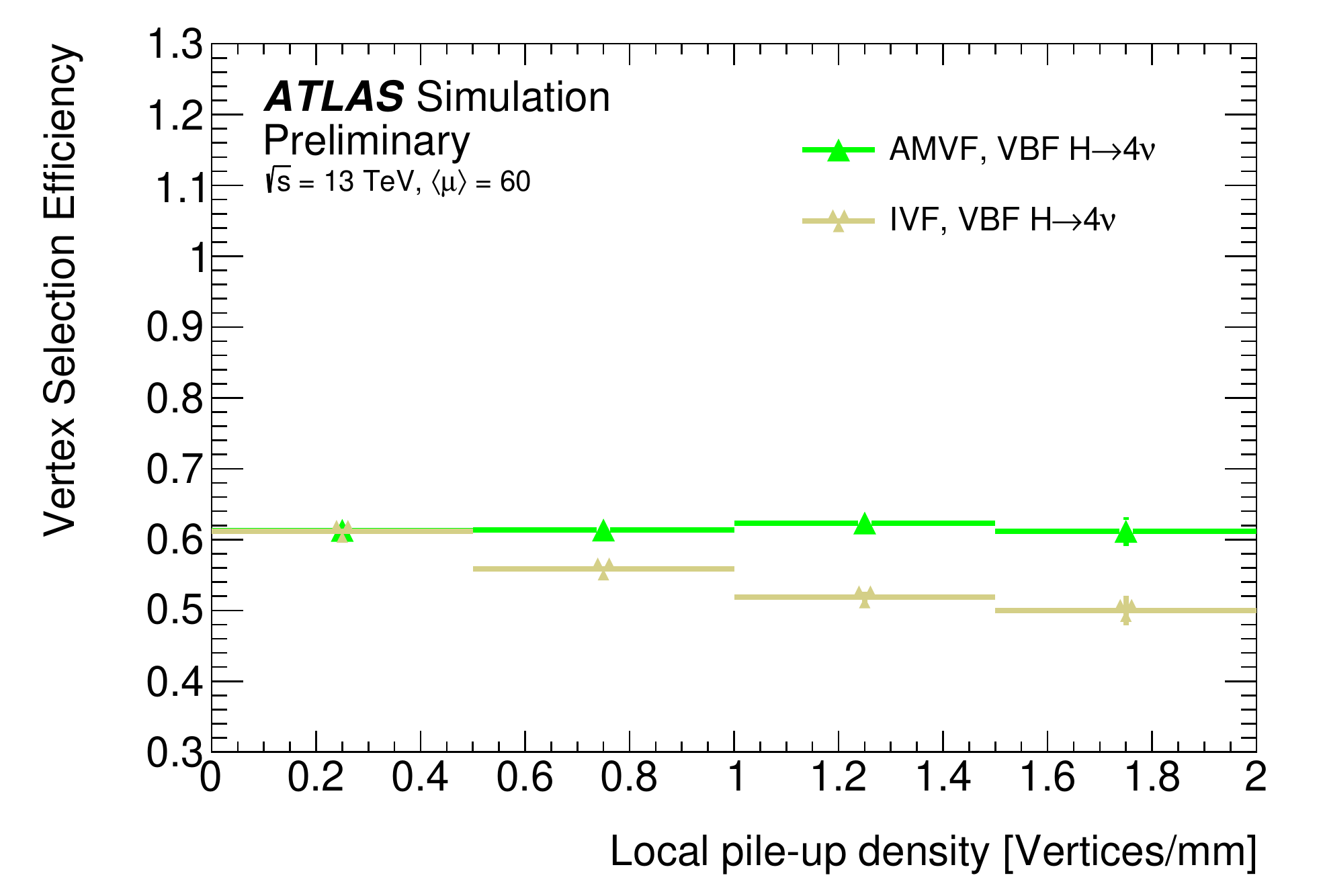}
    }
  \caption{Comparison of AMVF and IVF HS vertex selection efficiencies, as a function of local pile-up density. Figure \protect\subref*{fig:ttbarIdEff} (Figure \protect\subref*{fig:hinvIdEff}) shows performance for simulated $t\bar{t}$ ($\mathrm{VBF}\, H\to4\nu$) \protect\cite{ATLAS:PubNote}.}
  \label{fig:idEfficiencies}
\end{figure}
Reconstruction efficiency is defined as the fraction of events where the HS interaction is reconstructed within 0.1 mm of its true position. Selection efficiency is defined as the fraction of events where the HS is reconstructed and satisfies the $\Sigma p_T^2$ criterion. The reconstruction efficiency is shown in Figure \ref{fig:recoEfficiencies} and the selection efficiency is shown in Figure \ref{fig:idEfficiencies}. For $t\bar{t}$, the AMVF improves upon the already high reconstruction efficiency at all pile-up densities. In the case of $\mathrm{VBF}\, H\to 4\nu$, the drop in efficiency seen at high pile-up densities is about half that seen with the IVF. With regards to selection efficiency, the AMVF sees little to no pile-up dependence in either $t\bar{t}$ or $\mathrm{VBF}\, H\to 4\nu$.

\begin{figure}[!htb]
  \centering
  \subfloat[$t\bar{t}$]{
  	\label{fig:ttbarResR}
	\includegraphics[width=\imageSize]{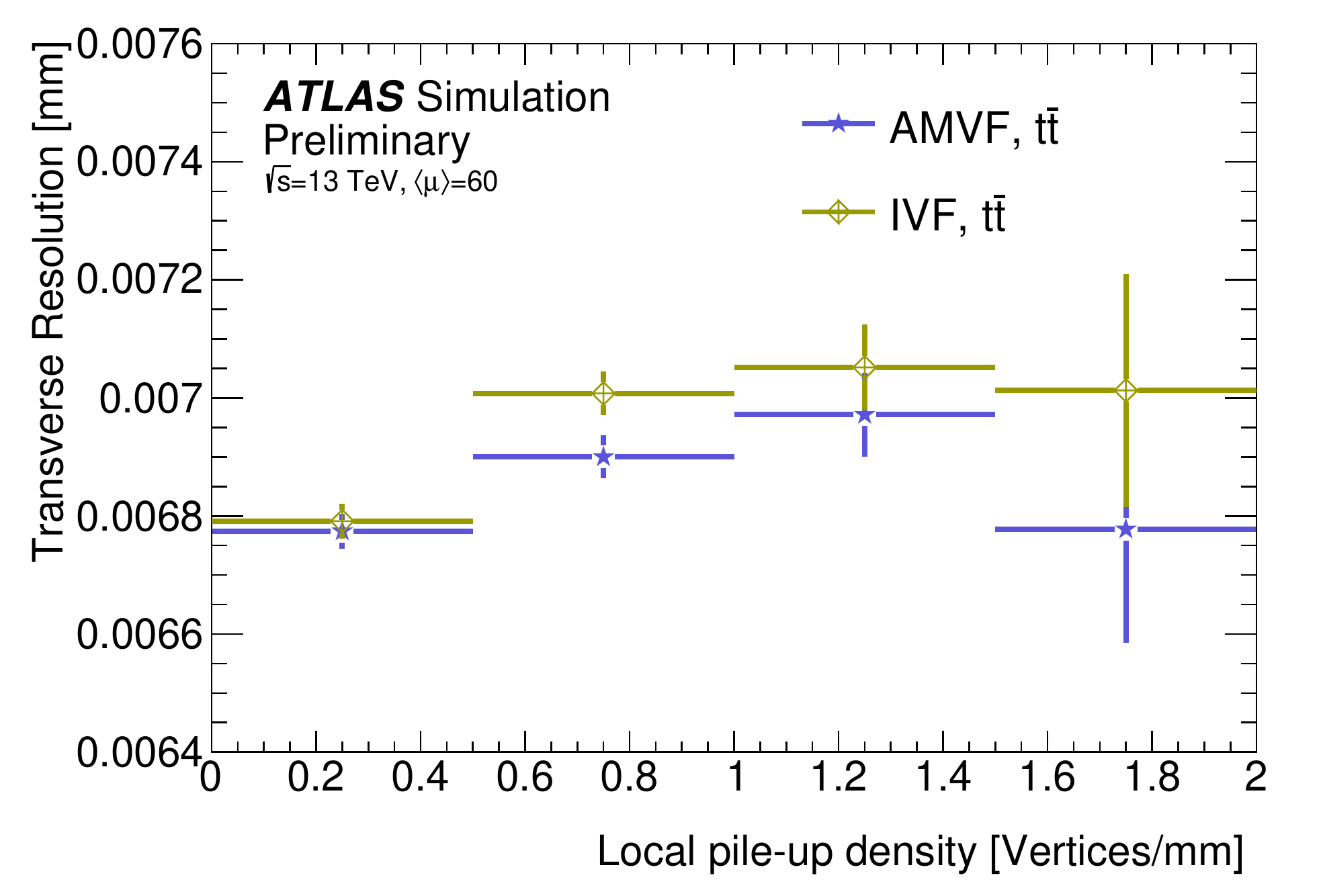}
	}
	\qquad 
  \subfloat[$\mathrm{VBF}\, H\to4\nu$]{
  	\label{fig:hinvResR}
  	\includegraphics[width=\imageSize]{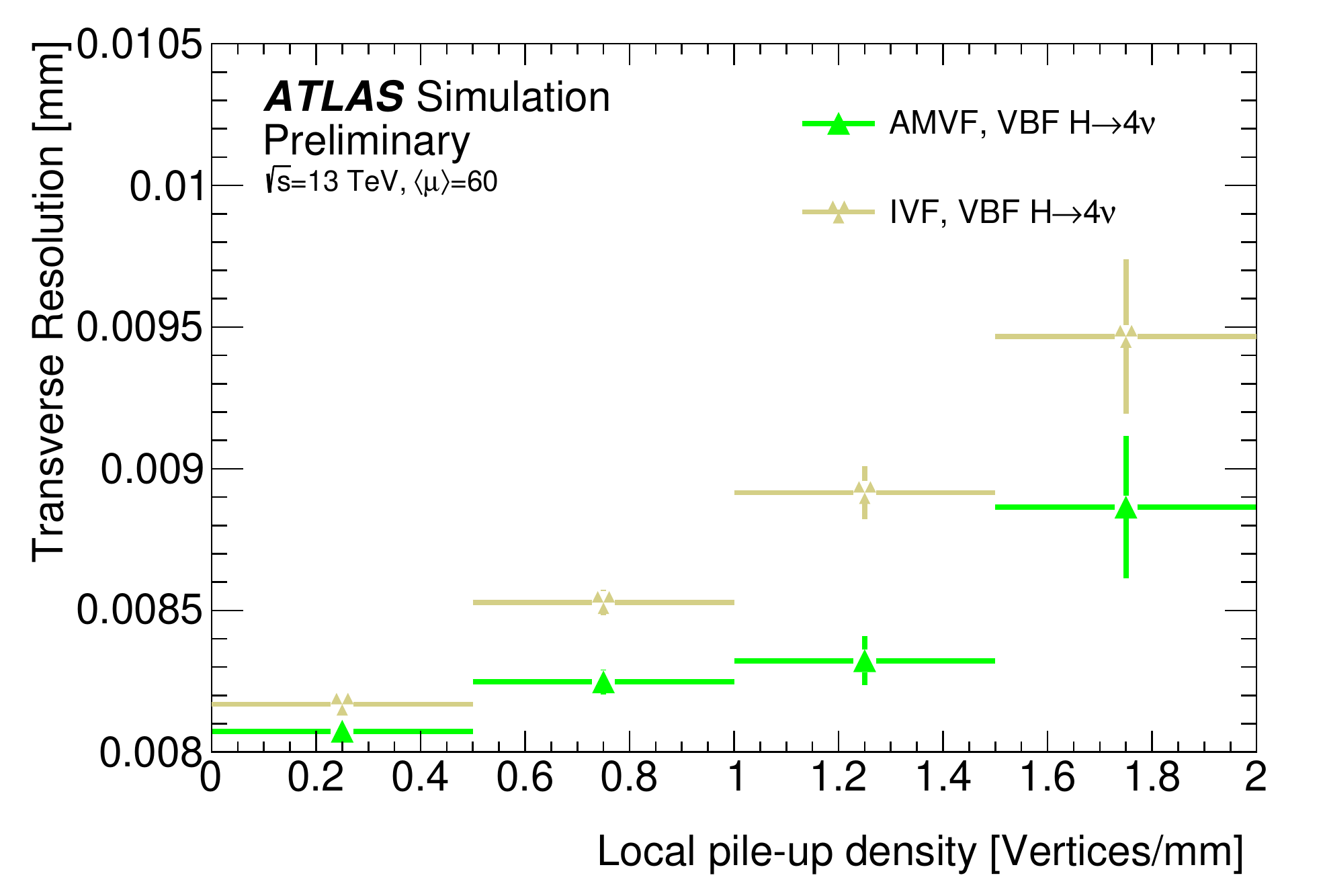}
  	}
  \caption{Comparison of AMVF and IVF HS transverse resolution, as a function of local pile-up density. Figure \protect\subref*{fig:ttbarResR} (Figure \protect\subref*{fig:hinvResR}) shows performance for simulated $t\bar{t}$ ($\mathrm{VBF}\, H\to4\nu$). The resolution is the difference in position between the generator level information and reconstructed vertex position, averaged over all events \protect\cite{ATLAS:PubNote}.}
  \label{fig:ttbarResolutions}
\end{figure}

\begin{figure}[!htb]
  \centering
  \subfloat[$t\bar{t}$]{
  	\label{fig:ttbarResZ}
  	\includegraphics[width=\imageSize]{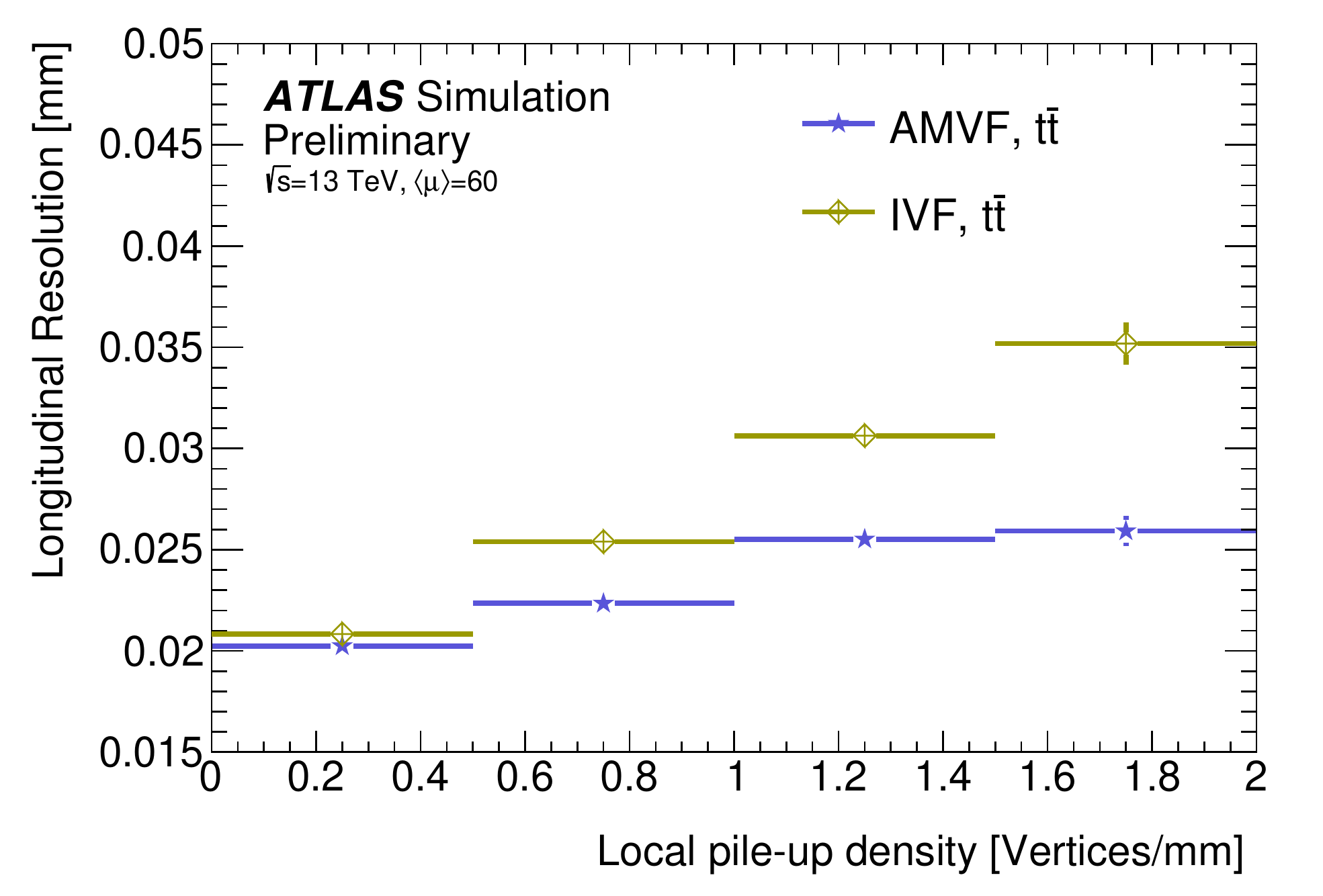}
  }
	\qquad 
  \subfloat[$\mathrm{VBF}\, H\to4\nu$]{
  \label{fig:hinvResZ}
  	\includegraphics[width=\imageSize]{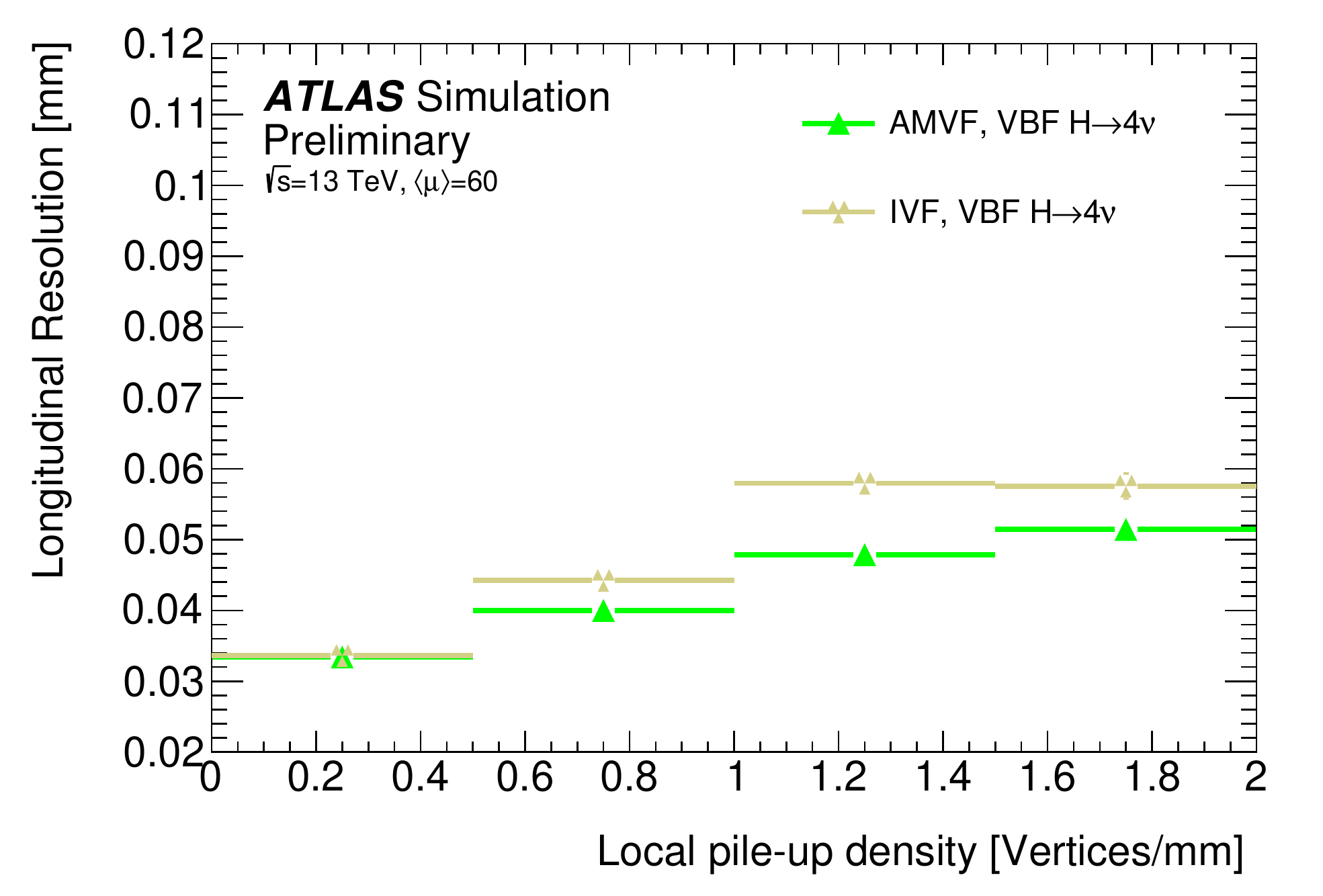}
  }
  \caption{Comparison of AMVF and IVF HS longitudinal resolution, as a function of local pile-up density. Figure \protect\subref*{fig:ttbarResZ} (Figure \protect\subref*{fig:hinvResZ}) shows performance for simulated $t\bar{t}$ ($\mathrm{VBF}\, H\to4\nu$). The resolution is the difference in position between the generator level information and reconstructed vertex position, averaged over all events \protect\cite{ATLAS:PubNote}.}
  \label{fig:hinvResolutions}
\end{figure}

Figures \ref{fig:ttbarResolutions} and \ref{fig:hinvResolutions} show the radial and longitudinal spatial resolutions calculated as the difference between the true and reconstructed vertex positions, averaged over all events in the sample. The radial resolution is largely constrained by the beam spot, so only minor improvements are seen, at the sub-micron level. In the $z$ direction the AMVF accomplishes up to a 20\% (10\%) better resolution for $t\bar{t}$ ($\mathrm{VBF}\, H\to4\nu$). 

\label{sec:HardScatterSpatialResolution}

\label{sec:HardScatterTrackEfficiencyAndPurity}
\begin{figure}[!htb]
  \centering
  \subfloat[$t\bar{t}$]{
  	\label{fig:ttbarHSratio}
  	\includegraphics[width=\imageSize]{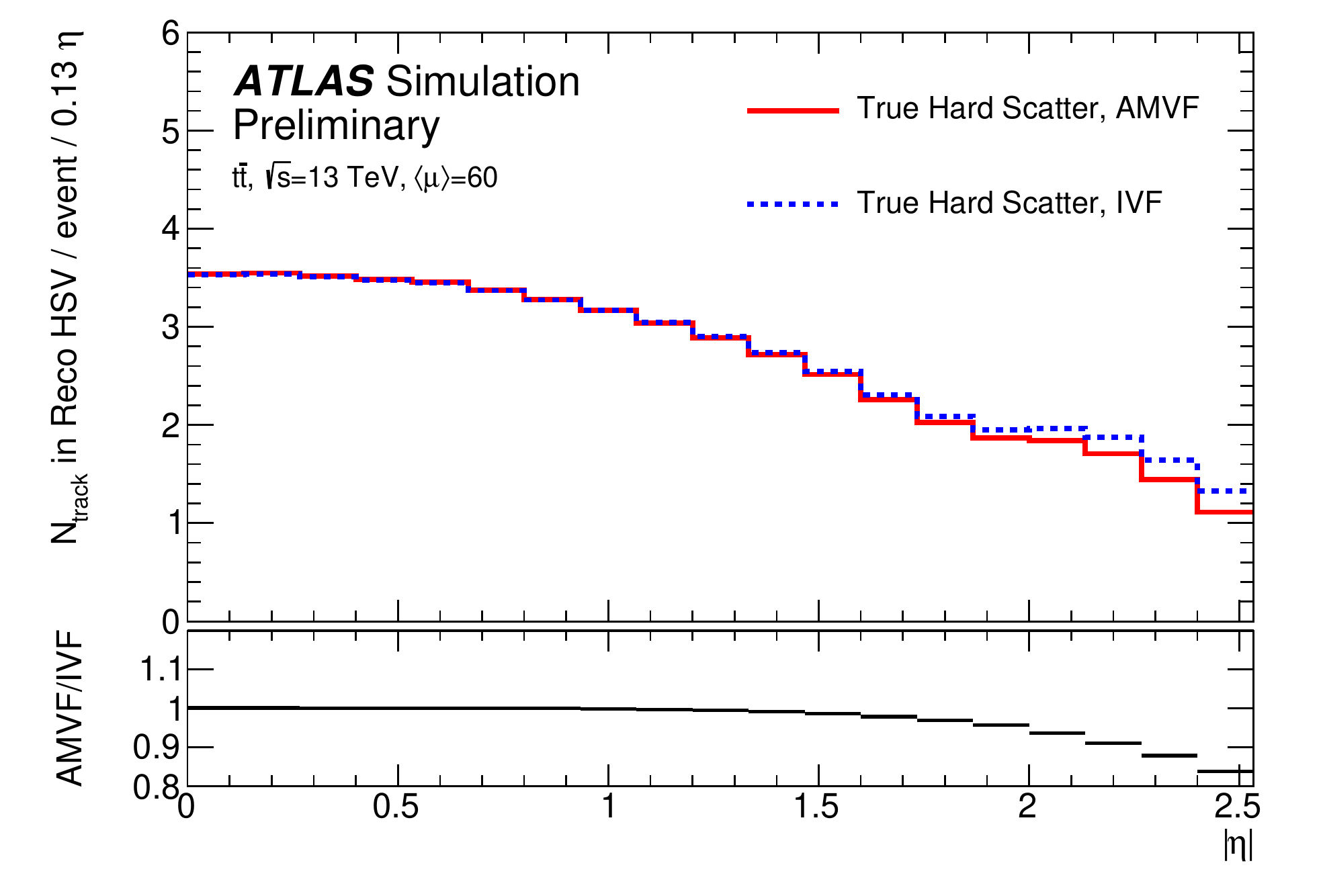}
  }
	\qquad 
  \subfloat[$\mathrm{VBF}\, H\to4\nu$]{
  	\label{fig:hinvHSratio}
  	\includegraphics[width=\imageSize]{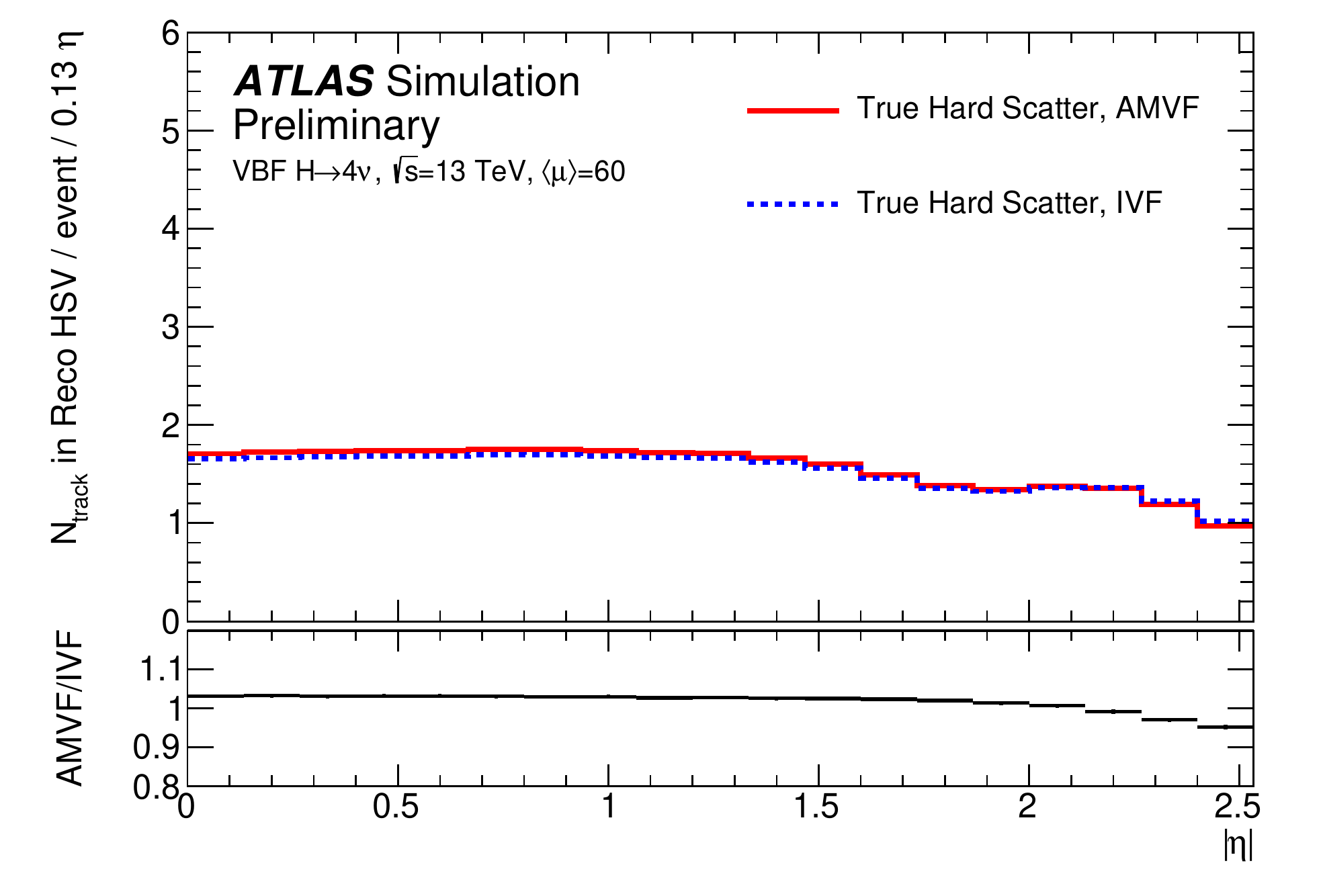}
  }
  \caption{Comparison of AMVF and IVF track efficiency as a function of $|\eta|$, for Figure \protect\subref*{fig:ttbarHSratio} (Figure \protect\subref*{fig:hinvHSratio}) simulated $t\bar{t}$ ($\mathrm{VBF}\, H\to4\nu$). The plots show the number of correctly associated compatible tracks per $\eta$ bin originating from the true HS vertex \protect\cite{ATLAS:PubNote}.}
  \label{fig:hsRatio}
\end{figure}

\begin{figure}[!htb]
  \centering
  \subfloat[$t\bar{t}$]{
  	\label{fig:ttbarPUratio}
  	\includegraphics[width=\imageSize]{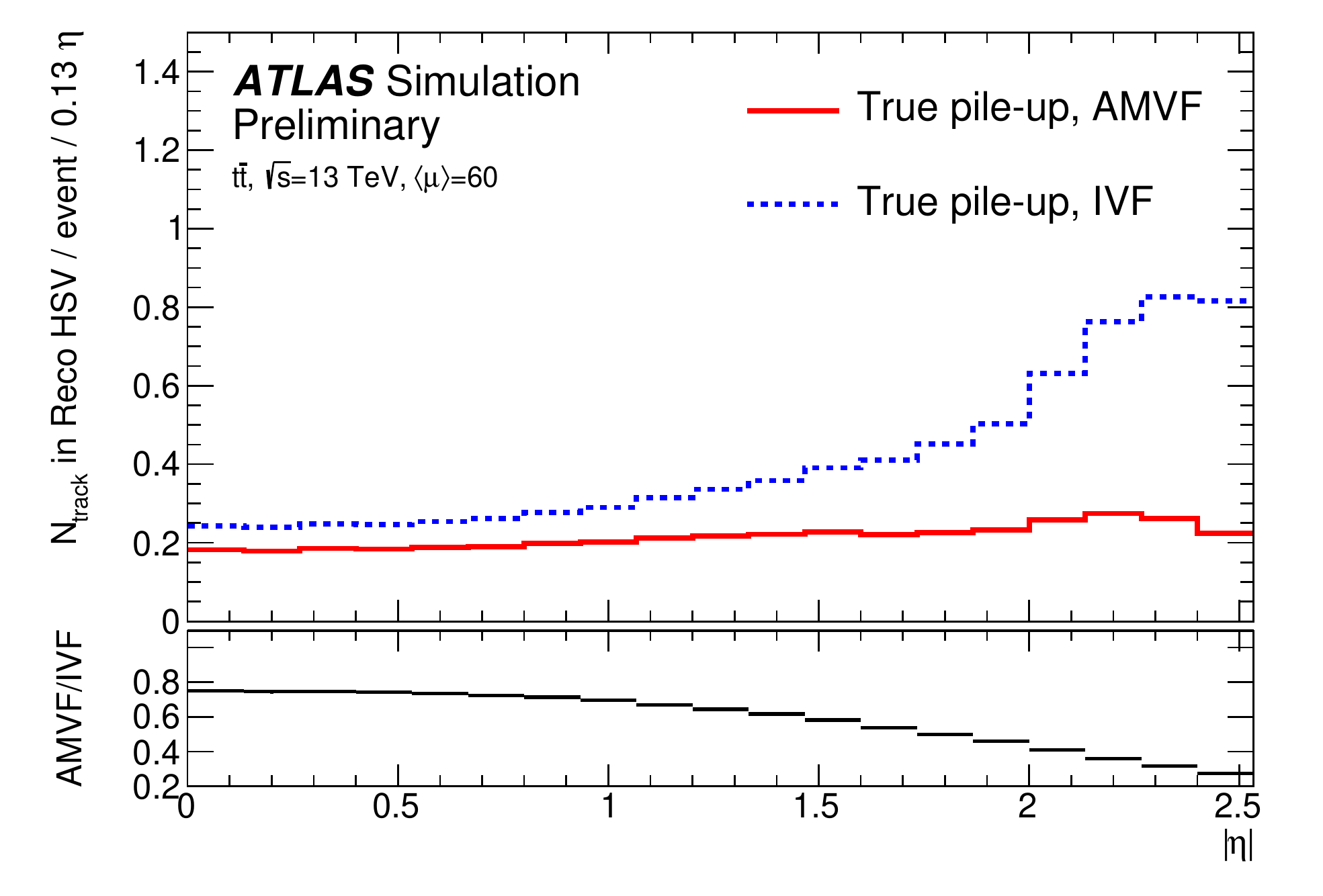}
  	}
	\qquad 
  \subfloat[$\mathrm{VBF}\, H\to4\nu$]{
	  \label{fig:hinvPUratio}
  	\includegraphics[width=\imageSize]{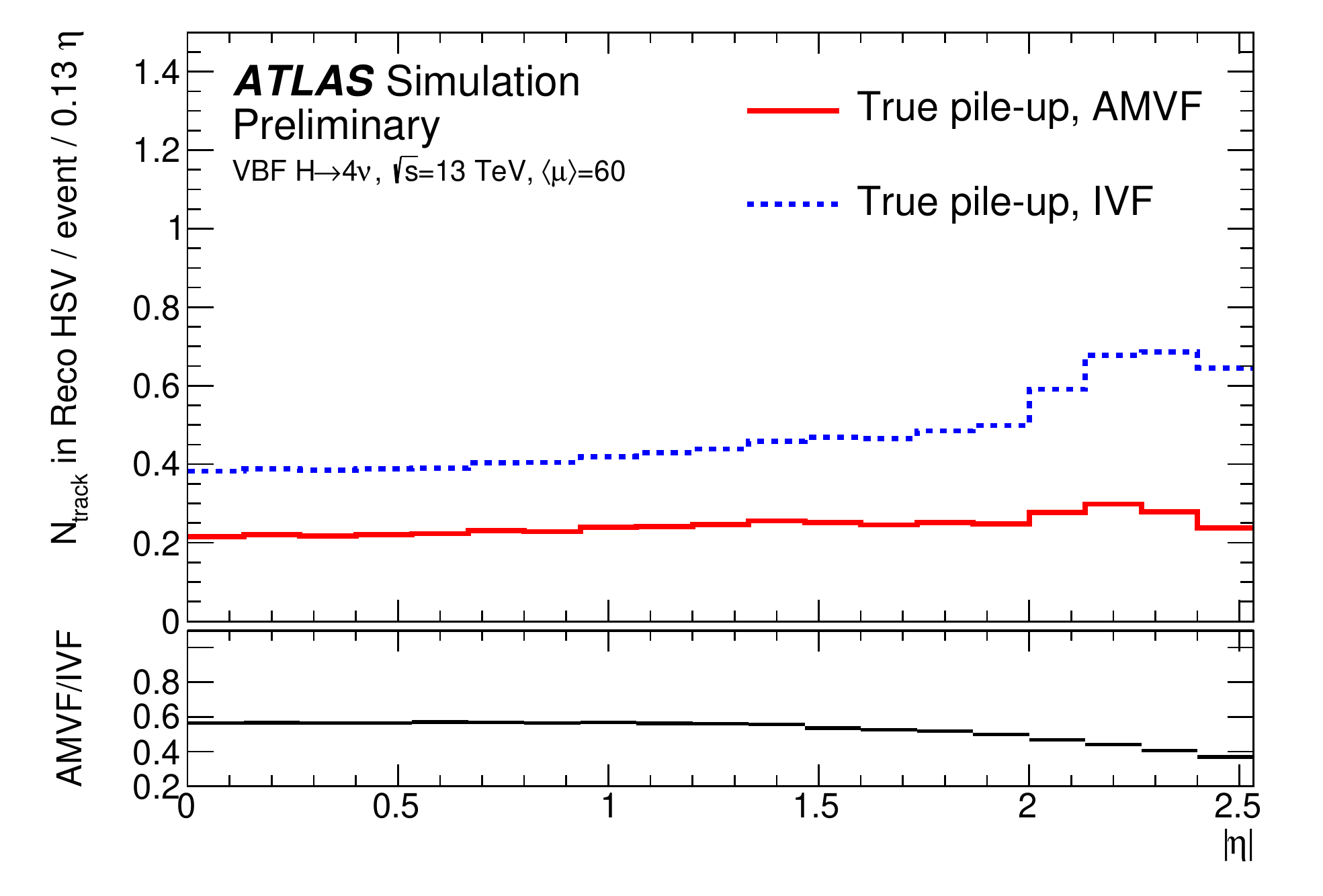}
  }
  \caption{Comparison of AMVF and IVF track contamination as a function of $|\eta|$, for Figure \protect\subref*{fig:ttbarPUratio} (Figure \protect\subref*{fig:hinvPUratio}) simulated $t\bar{t}$ ($\mathrm{VBF}\, H\to4\nu$). The plots show the number of compatible tracks per $\eta$ bin originating from true pile-up interactions and incorrectly associated to the HS vertex \protect\cite{ATLAS:PubNote}.}
  \label{fig:puRatio}
\end{figure}
Figure \ref{fig:hsRatio} shows the number of compatible ($\chi^2\leqslant 9$) tracks correctly assigned to the hard-scatter vertex as a function of absolute pseudorapidity, $|\eta|$, whereas Figure \ref{fig:puRatio} shows the number of incorrectly assigned tracks originating from pile-up interactions. For $t\bar{t}$, the AMVF and IVF demonstrate near identical performance in the central region ($|\eta|<1.25$). Larger $|\eta|$ tracks tend to have larger uncertainties, meaning they are compatible to a larger number of vertices. The IVF tends to assign these tracks to the first compatible vertex, which is usually the HS vertex, whereas the AMVF assigns these tracks to the vertex they are most compatible with. This is further seen in Figure \ref{fig:puRatio} where the AMVF sees a much lower (by 25-70\%) rate of pile-up track contamination across all $|\eta|$, with the largest difference at high $|\eta|$.

\vspace{\lessSpace}\subsection{Pile-up vertex reconstruction performance}
\begin{figure}[!htb]
  \centering
  \subfloat[Pile-up efficiency]{
  	\label{fig:puEfficiency}
  	\includegraphics[width=0.37\textwidth]{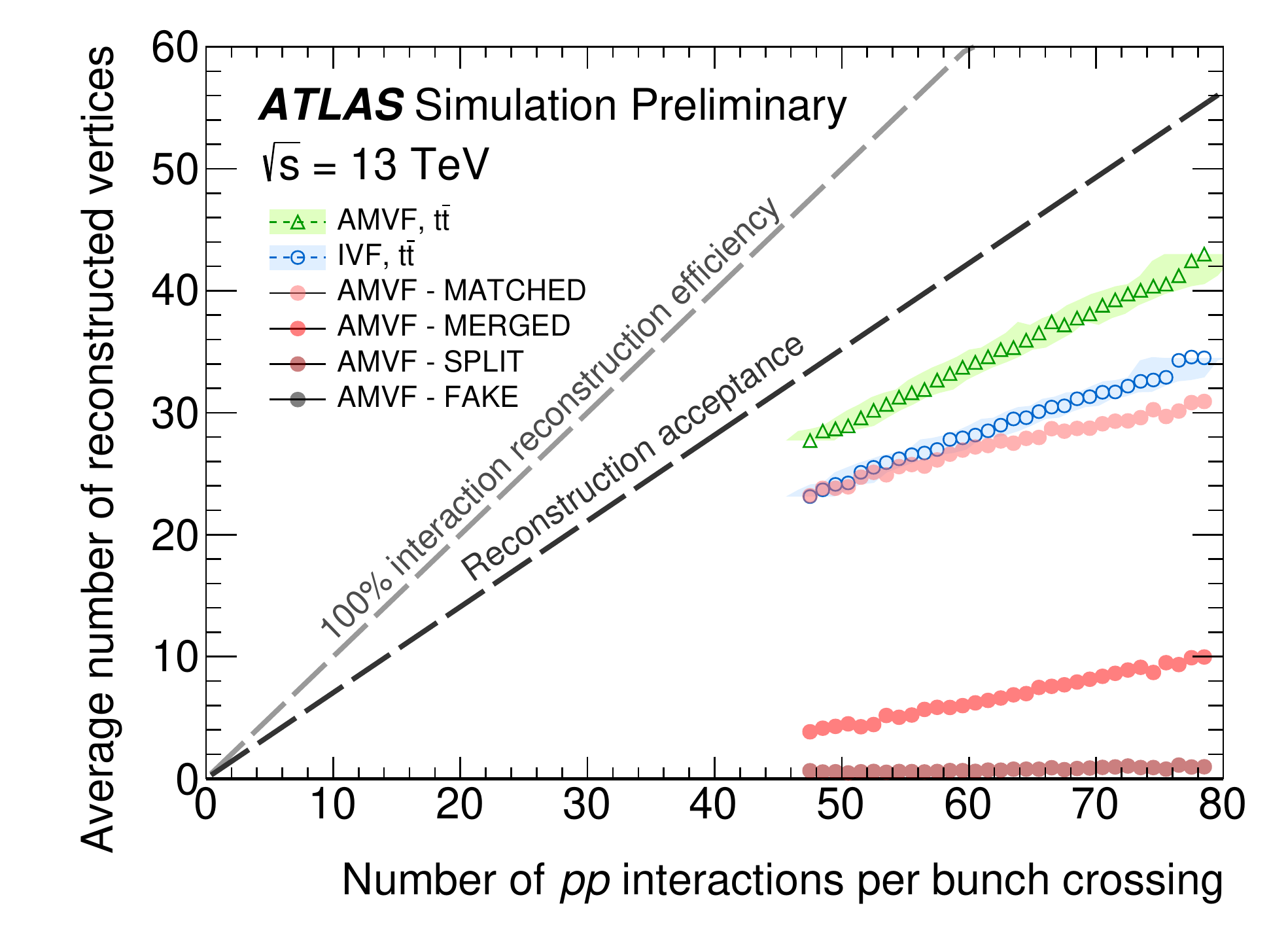}
  }
	\qquad 
  \subfloat[Vertex separation]{
  	\label{fig:puSeparation}
  	\includegraphics[width=\imageSize]{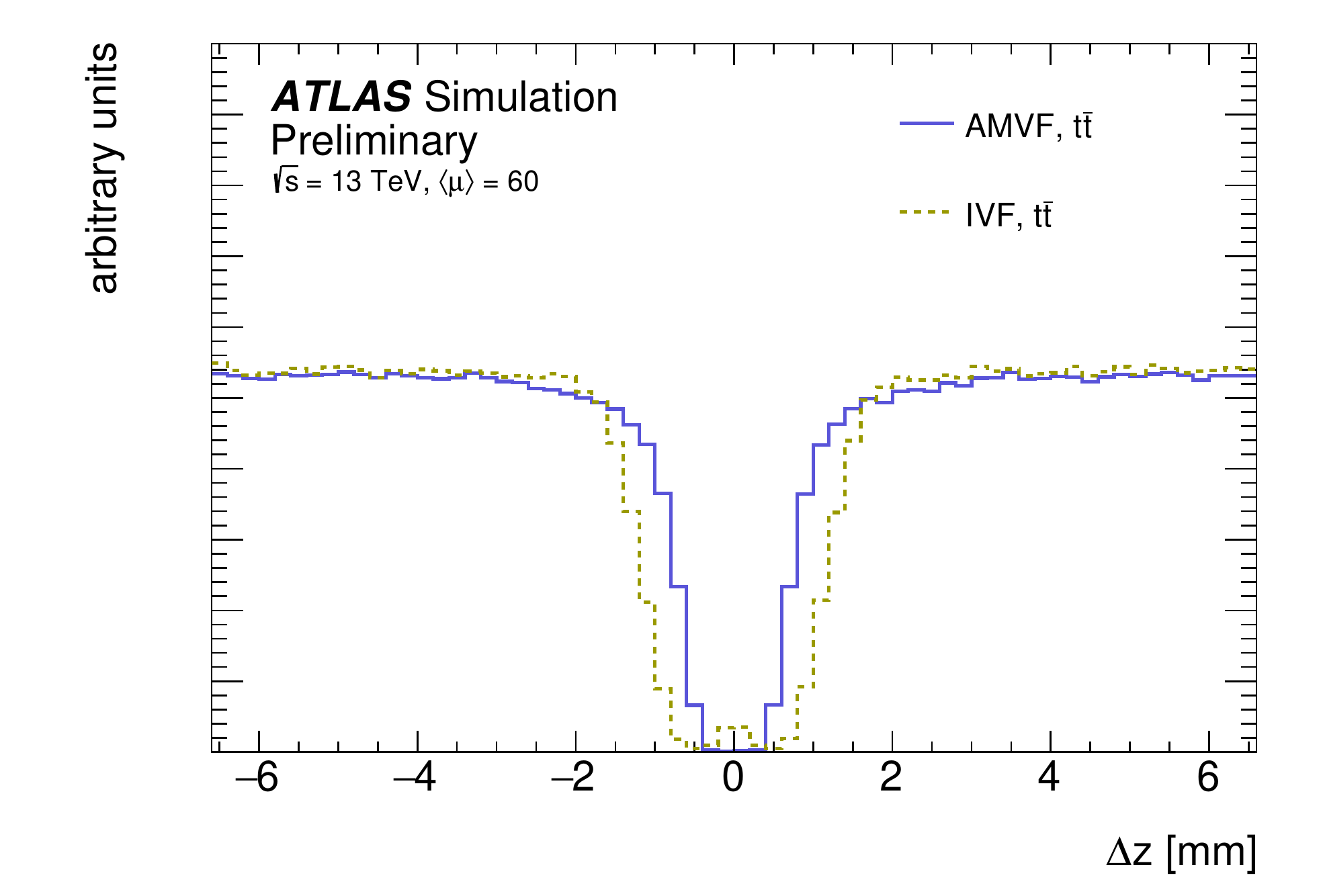}
  }
  \caption{Comparison of pile-up vertex reconstruction performance. Figure \protect\subref*{fig:puEfficiency} shows average number of vertices reconstructed as a function of the mean number of $pp$ interactions per bunch crossing, in simulated $t\bar{t}$ events. Figure \protect\subref*{fig:puSeparation} shows the distribution of the longitudinal separation between nearby reconstructed primary vertices in simulated $t\bar{t}$ events, for the IVF and the AMVF \protect\cite{ATLAS:PubNote}.}
  \label{fig:puPerformance}
\end{figure}
Figure \subref*{fig:puEfficiency} shows the average number of reconstructed vertices as a function of the true number of $pp$ collisions per bunch crossing, for the IVF and the AMVF. The upper dashed line shows the limit of 100\% efficiency, while the lower dashed line shows what would be achievable if every vertex  with tracks passing the quality selections were to be individually reconstructed. The performance of the AMVF is split into the different grades CLEAN/MATCHED, MERGED, SPLIT and FAKE. At high $\mu$, the AMVF recovers 35--50\% of the reconstructable primary interactions that the IVF is unable to reconstruct.

Figure \subref*{fig:puSeparation} shows the longitudinal difference in coordinates of all reconstructed vertices in each event, normalised to unity. The AMVF has a reduced tendency to merge closely spaced vertices, as shown by the narrower well, with adjacenct vertices separated by a millimetre or less. The depletion around $\Delta z = 0$~mm is due to merging. The AMVF is better able to resolve vertices with smaller separations, resulting in more closely spaced reconstructed vertices. The small excess seen for the IVF at $\Delta z$ = 0~mm is due to split vertices being reconstructed at the same position as their parent vertex, a behaviour that is prevented for the AMVF.

 






\vspace{\lessSpace}
\section{Conclusions}
The results presented in these proceedings show that the adaptive multi-vertex finder together with the Gaussian track density seed finder outperform the previous ATLAS vertex reconstruction strategy. The largest performance improvements are seen in higher pile-up environments. Such improvements are required for optimal physics performance in ATLAS Run 3 and beyond. 

Future work will aim to further exploit information available to the primary vertex finder from the new analytic seed-finding method, as well as to eliminate further sources of inefficiency within the AMVF. Computational efficiency has not been assessed nor improved thus far, though this will become a priority in future.


\vspace{-12pt}
\Acknowledgements
With thanks to D.~Casper, V.~Cairo, M.~Danninger, G.~R.~Lee, N.~Pettersson.

\vspace{\lessSpace}
\bibliography{eprint.bib}{}
\bibliographystyle{ieeetr}






\end{document}